# High-throughput screening of heterogeneous transition metal dual-atom catalysts by synergistic effect for nitrate reduction to ammonia

Zheng Shu[1], Hongfei Chen[1], Xing Liu[1], Huaxian Jia[2], Hejin Yan[1], Yongqing Cai[1,*]

[1]Joint Key Laboratory of the Ministry of Education, Institute of Applied Physics and Materials Engineering, University of Macau, Taipa, Macau, China

[2]Beijing National Laboratory for Condensed Matter Physics and Institute of Physics, Chinese Academy of Sciences, Beijing 100190, China

*Corresponding author: yongqingcai@um.edu.mo

**Nitrate reduction to ammonia has attracted much attention for nitrate ($NO_3^-$) removal and ammonia ($NH_3$) production. Identifying promising catalyst for active nitrate electroreduction reaction ($NO_3$RR) is critical to realize efficient upscaling synthesis of $NH_3$ under low-temperature condition. For this purpose, by means of spin-polarized first-principles calculations, the $NO_3$RR performance on a series of graphitic carbon nitride (g-CN) supported double-atom catalysts (denoted as $M_1M_2$@g-CN) are systematically investigated. The synergistic effect of heterogeneous dual-metal sites can bring out tunable activity and selectivity for $NO_3$RR. Amongst 21 candidates examined, FeMo@g-CN and CrMo@g-CN possess a high performance with low limiting potentials of -0.34 and -0.39 V, respectively. The activities can be attributed to a synergistic effect of the $M_1M_2$ dimer *d* orbitals coupling with the anti-bonding orbital of $NO_3^-$. The dissociation of deposited FeMo and CrMo dimers into two separated monomers is proved to be difficult, ensuring the kinetic stability of $M_1M_2$@g-CN. Furthermore, the dual-metal decorated on g-CN significantly reduces the bandgap of g-CN and broadens the adsorption window of visible light, implying its great promise for photocatalysis. This work opens a new avenue for future theoretical and experimental design related to $NO_3$RR photo-/electrocatalysts.**

Water resources are currently suffering from the ever-increasing accumulation of nitrate ($NO_3^-$) due to human activities such as agriculture and industry.[1,2] As a common byproduct of industrialization and agricultural activities, $NO_3^-$ pollutant becomes one of the foremost global risk factors for diseases.[3,4] Efficient removal of nitrate is a critical issue for renewable development. Fortunately, the nature has its own way to convert $NO_3^-$ via anaerobic ammonium oxidation of bacteria using $NO_3^-$ as the electron acceptor.[5] Greatly inspired by the natural process, artificial removal of $NO_3^-$ while concomitant conversion it into non-toxic or value-added compounds, such as nitrogen ($N_2$) and ammonia ($NH_3$), is the goal of researchers for global nitrogen cycle.[6] In particular, efficiently transforming $NO_3^-$ to $NH_3$ is of significant importance to maintain the balance of natural nitrogen cycle and remediate the crisis of energy.[7,8]

$NH_3$ is a key ingredient for the production of fertilizers, pharmaceutical and light industries.[9,10] Traditional way of the $NH_3$ production the Haber-Bosch approach relies on extremely harsh conditions (200-250 bar, 400-500 °C) and abundant hydrogen consumption via burning huge non-renewable fuels.[11] Electrochemical nitrogen reduction reaction (NRR) is a fascinating strategy for direct electroreduction of $N_2$ to $NH_3$ under ambient conditions.[12-14] Nonetheless, the massive production via NRR is still a long way to go due to the insolubility and the inert N≡N bond of nitrogen molecule.[15] As an alternative method, electrochemical reduction of nitrate-to-ammonia ($NO_3RR$) is a more sustainable pathway for $NO_3^-$ remediation and concomitant $NH_3$ production under room-temperature.[16,17] In comparison to NRR, the $NO_3RR$ is more promising for $NH_3$ production owing to a much easier cracking of N=O bond (204 kJ $mol^{-1}$) than the inert N≡N bond (941 kJ $mol^{-1}$).[18] Therefore, electrochemical $NO_3RR$ is a "two birds with one stone" avenue for addressing nitrate-based environmental problem while simultaneously producing $NH_3$ under low-temperature. The total nitrate reduction ($NO_3^- + 9H^+ + 8e^- \rightarrow NH_3 + 3H_2O$) is an eight-electron electron-proton transfer reaction, which can be decomposed into a series of deoxygenation and hydrogenation processes in an acidic environment.[17,19] Consequently, elucidating the mechanism for $NO_3RR$ pathways and designing highly efficient electrocatalysts are critically important. Until now, screening of active, selective and stable electrocatalysts with controllable reaction pathway and selectivity for $NO_3RR$ remains a great challenge.

Copper and copper-based derivatives have proven to be active electrocatalysts for $NO_3RR$.[19-22] However, the adsorption of $NO_3^-$ on Cu (111) surface is an endothermic process, which hinders the $NO_3RR$ activity. In contrast with bulk surface, single-atom catalysts (SACs) have been in the spotlight with superior catalytic activity for $NO_3RR$.[23-29] The maximum atom-utilization efficiencies of SACs is adequate, but only one active site of it brings out difficulties to break the linear scaling relations for complicated reactions.[30] Recently, double-atom catalysts (DACs) as an extension of SACs have brought about great success in many catalytic reactions.[31] Owing to the synergic effect and flexible controllability of dual-metal active sites, DACs can expand the active configurational space, desired for multistep reactions like $NO_3RR$.[32-36] Sun et al. reveled that PdCu dual-atom anchored on phosphorene plays a synergistic role in $NO_3RR$,[35] and $PdCuP_4$/CS cathode with special coordination structure exhibits excellent $NO_3^-$ removal rate. Rehman et al. found that Cr dimer supported on the 2D expanded phthalocyanine exhibits a good performance for the conversion of $NO_3^-$ to $NH_3$ with the limiting potential of -0.02 V.[30] Lv and co-workers presented a concept-new heterogeneous bilayer single-atom catalysts (BSACs) to boost the activity and selectivity of $NO_3RR$ by controllable electrochemical process.[36] However, the design of highly active $NO_3RR$ electrocatalysts is still in the early stage, and in particular, the catalytic mechanism of heterogeneous dual-metals DACs for $NO_3RR$ remains unknown.

In this work, a series of heterogeneous bimetallic DACs embedded on the graphitic carbon nitride (g-CN) is studied as the potential electrocatalysts for $NO_3RR$ by high-throughput *in silico* first-principles calculations based on density functional theory (DFT). Via examining 21 candidates of $M_1M_2$@g-CN, 17 heterogeneous DACs which meet the thermodynamic stability criteria are screened out for further investigation. By comparing the limiting potential and selectivity of above DACs, FeMo@g-CN is highlighted for $NO_3RR$ with a limiting potential of -0.34 V. Meanwhile, CrMo@g-CN and CrRu@g-CN are also promising DACs for $NO_3RR$ with limiting potentials of -0.39 and -0.49 V, and ZrMo@g-CN is promising for hydrogen evolution reaction (HER) with an appropriate free energy of hydrogen adsorption ($\Delta G_{H*}$) of nearly zero. Moreover, these DACs possess suitable bandgaps and band edge for photocatalysis. In light of the electronic analysis, the $NO_3RR$ activation originates from synergistic effect of the $M_1M_2$ dimer *d* orbitals coupling with the anti-bonding orbital of $NO_3^-$. In the regard, we suggest that embedding heterogeneous dual-metals on g-CN is a flexible and emerging selection for $NO_3RR$ and worthy of further experimental verification.

## Results

**Structural stability and $NO_3^-$ adsorption of the $M_1M_2$@g-CN catalysts.** The porous g-CN monolayer belongs to the hexagonal crystal system with the space group of P-6m2, and its primitive cell contains 12 atoms. The relaxed lattice constant of the lowest-energy structure is $a$ = 7.12 Å, which agrees well with previous works.[33] The graphite-like CN layer has been experimentally synthesized in a large scale through a simple solvothermal technique.[37] Compared with $g\text{-}C_3N_4$ monolayer, larger N-edged cavities in g-CN monolayer are more appropriate to serve as a support for DACs by $N_3$-$M_1$-$M_2$-$N_3$ bonding.[33] Herein, the model of dual-metals embedded in the periodic cavities of g-CN monolayer is shown in **Figure 1**a.

We took a series of combinations of heterogeneous dual-metals into account, including Ti, Cr, Fe, Cu, Zr, Mo and Ru that have previously been used as SACs with desirable performance to catalyze $NO_3RR$.[8,23,24,28,29,38,39] Thus, a total of 21 types of $M_1M_2$@g-CN structures are constructed (two out of Ti, Cr, Fe, Cu, Zr, Mo, Ru are taken in combination to form TiCr, TiFe, TiCu, TiZr, TiMo, TiRu, CrFe, CrCu, CrZr, CrMo, CrRu, etc. heteronuclear bimetallic catalysts) and the corresponding optimized geometrical configurations of them are shown in Figure S1 (Supporting Information). Since Cu-Zr dimer can't form a planar structure on the g-CN, it is ruled out in the next discussion. The out-of-plane displacement ($h_{M1\text{-}M2}$) and bond length ($d_{M1\text{-}M2}$) of dual-metals are shown in Figure S2. As we can see, the out-of-plane displacement of TiCr and TiMo dimers are nearly zero, making them largely locating within the plane of g-CN monolayer. The TiZr dimer has the highest relative outward movement, which may make it difficult to adsorb $NO_3^-$. The motivation of designing such DAC systems is that the bimetallic systems contain richer electronic states to possibly activate multiple intermediates in $NO_3RR$ while stabilize their adsorption at the g-CN support.

For DACs, possible diffusion and aggregation of dual-metals are severe issues reducing the stability and endurance of catalysts. To evaluate the energetics and kinetic stability of the DACs, the formation energy ($E_{form}$) of dual-metals on g-CN were examined (Table S1, Supporting Information) according to the difference of binding energy ($E_b$) and cohesive energy ($E_{coh}$): $E_{form} = E_b – E_{coh} = (E_{M1M2@g\text{-}CN} – E_{g\text{-}CN} - E_{M1\text{-}bulk}/n - E_{M2\text{-}bulk}/m)/2$, ,[32,33] where $E_{M1M2@g\text{-}CN}$, $E_{g\text{-}CN}$ , and $E_{M1\text{-}bulk}$ ($E_{M2\text{-}bulk}$) are the total energies of $M_1M_2$@g-CN, g-CN, and $M_1$ ($M_2$) metal bulk, respectively, and $n$ and $m$ are the number of metal atoms in the unit cell of the bulk metal $M_1$ and $M_2$, respectively. Amongst the configurations, the $E_{form}$ ranges from +0.56 to -1.40 eV, a negative value indicating an energetically favorable adsorption against aggregating into clusters. As shown in **Figure 1**a, 17 potential candidates (TiCr, TiFe, TiCu, TiZr, TiMo, TiRu, CrFe, CrCu, CrZr, CrMo, CrRu, FeCu, FeZr, FeMo, ZrMo, ZrRu and MoRu DACs) with a negative $E_{form}$ are screened for further investigation. The exact values of $E_{form}$, charge transfers and bond lengths with adjacent $N$ atoms of dual-metals are summarized in Table S1 (Supporting Information). During $NO_3RR$ process, the empty d-orbitals of metal dimer can accept the electrons from $NO_3^-$ or other intermediates to strengthen the TM-O adsorption. The TM-O hybridization in turn spills some charges to occupy the anti-bonding orbital ($\pi^*$) of $NO_3^-$ and weaken the N=O bond (**Figure 1**b). For SACs, the TM active sites can accept the electrons of $NO_3^-$, and simultaneously donate d electrons to empty $\pi^*$ orbital according to the "acceptance-donation" mechanism. To maximize the activation potential, DACs can make use of d-orbitals of the bimetallic sites via "pull-pull" effect,[40] as shown in **Figure 1**b. It can also take advantage of the innate dual-site to increase the adjusted space.

The first step to catalyze the $NO_3RR$ reaction is the adsorption of $NO_3^-$ above $M_1M_2$@g-CN. The Gibbs free energies of $NO_3^-$ adsorption ($\Delta G_{*NO3}$) on these DACs are calculated and plotted in **Figure 1**c. Our results reveal that the $NO_3^-$ prefers to bonding with the dual-metals by adopting an O-bidentate configuration (two oxygen atoms adsorbed on different metal atoms, see Figure S3, Supporting Information), where $\Delta G_{*NO3}$ ranges from -1.47 to 1.61 eV. The oxygen coordination with catalytic sites of the electrode surface is important to "pull-pull" effect for subsequent reactions. Notably, the $\Delta G_{*NO3}$ of most $M_1M_2$@g-CN are negative except for TiZr@g-CN, FeZr@g-CN and ZrRu@g-CN, making them being excluded due to the poor adsorption of $NO_3^-$. Therefore, 14 out of 17 candidates that meet the criterion $\Delta G_{*NO3} < 0$ are selected as the potential $NO_3RR$ catalysts. The schematic setup of experimental electrochemical $NO_3RR$ at the $M_1M_2$@g-CN cathode in the electrolytic cell ($RuO_2$ as the cathode) is illustrated in **Figure 1**d, together with the oxygen evolution reaction at the anode.

**$NO_3RR$ paths and performance.** The $NO_3RR$ involves a complicated multistep electron-proton transfer process. In general, there are three kinds of $NO_3RR$ pathways on metal surface or dual-atoms in the literature reports.[19-21,41-43] One is a series of deoxidation and hydrogenation elementary reactions, which form the intermediate *N: $NO_3^- \rightarrow *NO_3 \rightarrow *NO_2 \rightarrow *NO \rightarrow *N \rightarrow *NH \rightarrow *NH_2 \rightarrow *NH_3 \rightarrow NH_3$ (g). In other pathways, the first hydrogenation occurs at *NO, which are described as: $NO_3^-$ → $*NO_3$ → $*NO_2$ → *NO → *NOH → *NHOH → *NH → $*NH_2$ → $*NH_3$ → $NH_3$ (g), and $NO_3^-$ → $*NO_3$ → $*NO_2$ → *NO → *NOH → *NHOH → $*NH_2OH$ → $*NH_2$ → $*NH_3$ → $NH_3$ (g), respectively. The main difference of these three pathways is the sequence of deoxidation and hydrogenation during the $NO_3RR$ process. However, the hydrogenation of $*NO_x$ to form $*NO_xH$ is not included in these three pathways. Certainly, there are many other pathways for $NO_3RR$,[8,24-26,29] e.g., $NO_3^-$ → $*NO_3$ → $*NO_3H$ → $*NO_2$ → $*NO_2H$ → *NO → *NOH → *N → *NH → $*NH_2$ → $*NH_3$ → $NH_3$ (g). Based on above discussions, the three pathways for $NO_3RR$, namely Path I, II and III, are presented to investigate the nitrate-to-ammonia process in detail, which is shown in **Figure 2**a.

All potential $M_1M_2$@g-CN electrocatalysts were systematically examined by Path I, II and III, and their free energy evolutions with correction ($E_{ZPE}$-$TS$) for each elementary step are listed in Tables S2-S5 (Supporting Information). Based on DFT calculations, the true reaction pathway of $NO_3RR$ can be deduced by identifying the path with the lowest thermodynamic barrier ($\Delta G$ in each elementary step). **Figure 2**b summarizes the limiting potentials ($U_L$) for $NO_3RR$ to form $NH_3$ on 14 potential $M_1M_2$@g-CN DACs. It can be seen that FeMo@g-CN exhibits the best $U_L$ of -0.34 V. In addition, CrMo@g-CN and CrRu@g-CN also exhibit the desirable $U_L$ of -0.39 and -0.49 V, respectively. Furthermore, the free energy change of desorption of $NH_3$ ($\Delta G_{\mathbf{*NH3\text{-}des}}$), a critical indicator for reaction activity, is also summarized in Figure S4 (Supporting Information). It is seen that most catalysts have a moderate $\Delta G_{\mathbf{*NH3\text{-}des}}$ from -0.23 to 1.65 eV, implying a good sustainability for $NO_3RR$ process. On the other hand, hydrogen evolution reaction (HER) as the main competing side reaction can restrain the Faradaic efficiency (FE), and block the active site of $NO_3RR$. In some works, authors think that a catalyst with a stronger $\Delta G_{\mathbf{*NO3}}$ would be promising for the selectivity towards $NO_3RR$.[8,26] However, $U_L$ is the decisive factor for the whole reaction in electrocatalytic process.[36] Thus, the $U_L$ is more suitable to evaluate the selectivity of $NO_3RR$ and HER. To check if $NO_3RR$ is more energetically favorable than HER, we compared the limiting potentials, $U_L$ ($NO_3RR$) and $U_L$ (HER), of $NO_3RR$ and HER. The values of $\Delta G_{\mathbf{*H}}$ and the corresponding configurations are shown in **Figure 2**c and S5 (Supporting Information), respectively. As shown in **Figure 2**d, the region below the diagonal line represents the $NO_3RR$ selective, while the region above the diagonal line denotes the HER selective. It is revealed that half of the potential DACs (TiCr@g-CN, TiFe@g-CN, TiCu@g-CN, TiRu@g-CN, CrMo@g-CN, CrRu@g-CN and FeMo@g-CN) meet the selectivity criterion toward $NH_3$ during the whole reaction process. By the way, ZrMo@g-CN exhibits high efficient HER performance with the $\Delta G_{\mathbf{H*}}$ of nearly zero.

**Figure 3**a,b gives a panorama of $NO_3RR$ along the possible energy pathway on FeMo@g-CN and CrMo@g-CN as the representative catalysts to analyze the process of nitrate-to-ammonia, and corresponding reaction configurations are shown in **Figure 3**c,d. We can see that $NO_3^-$ is stably adsorbed on dual-metal site and then go through a series of exothermic process to form $*NO_2$ with the energy change of -0.75, -1.18, and

-0.92 eV for FeMo@g-CN and -0.79, -0.99, and -1.10 eV for CrMo@g-CN. In above process, one $H_2O$ molecule is released through $*NO_3H + H^+ + e^- \rightarrow *NO_2 + H_2O$. In the subsequent step, $*NO_2$ is hydrogenated by proton-electron coupling pair ($H^+ + e^-$) to form the intermediate $*NO_2H$. This process is a slightly endothermic step for FeMo@g-CN with a $\Delta G$ of 0.12 eV, but is exothermic step for CrMo@g-CN with a $\Delta G$ of -0.10 eV. Next, the second $H_2O$ molecule is released when $*NO_2H$ is attacked by ($H^+ + e^-$) pair to form *NO intermediate. The changes of Gibbs free energy are -2.10 and -1.79 eV for FeMo@g-CN and CrMo@g-CN, respectively. Subsequently, the hydrogenation step of *NO to *NOH is observed as the potential-determining step (PDS) with $\Delta G_{max}$ of 0.34 and 0.39 eV for FeMo@g-CN and CrMo@g-CN, respectively. In the following processes along Path I ($*NOH \rightarrow *N \rightarrow *NH \rightarrow *NH_2$), ($H^+ + e^-$) pair consecutively attacks a series of intermediates with downhill $\Delta G$ of -1.57, -0.63 and -0.71 eV for FeMo@g-CN and -1.66, -0.48 and -0.72 eV for CrMo@g-CN. Finally, $*NH_2$ intermediate is hydrogenated to form $*NH_3$, then $NH_3$ escapes from the surface of catalyst by desorption process. The $\Delta G$ are 0.01 and 1.20 eV for FeMo@g-CN, and -0.14 and 1.19 eV for CrMo@g-CN in these hydrogenation and desorption processes respectively. Therefore, $NO_3^-$ is successfully reduced to $NH_3$ on FeMo@g-CN and CrMo@g-CN through Path I. It is noticed that the energy barriers of $*NOH \rightarrow *NHOH$ (0.31 eV for FeMo@g-CN and -0.09 eV for CrMo@g-CN) in Path II are much larger than those of $*NOH \rightarrow *N$ along Path I, suggesting that Path II is hard to happen compared to Path I. The reaction panoramas of other catalysts are shown in Figures S6–S11 (Supporting Information). Taking these DFT calculations into account, Path I is energetically favorable than Path II and III on most of $M_1M_2$@g-CN, except for CrZr@g-CN. $NO_3RR$ on CrZr@g-CN prefers to adopt Path III and its corresponding PDS occurs in the step of $*NH_2OH \rightarrow *NH_2$ with a $\Delta G_{max}$ of 1.15 eV.

Another issue of $NO_3RR$ is the formation of byproducts, mainly toxic gases such as $NO_2$ and NO. The deoxidation during the reaction process suggests the formation of $NO_2$ and NO is infeasibility with the energy barriers of 2.26 and 2.61 eV for FeMo@g-CN, as shown in **Figure 3**a (2.29 and 2.55 eV for CrMo@g-CN shown in **Figure 3**b). Similar tendency occurs in other catalysts (Figures S6–S11, Supporting Information), implying the release of toxic gases during the reaction is almost impossible. In addition, the kinetic barriers for the release of $NO_2$, NO and $NH_3$ on four selective DACs, CrMo@g-CN, CrRu@g-CN, FeMo@g-CN and ZrMo@g-CN, are examined via CI-NEB method. As shown in Figure S12 (Supporting Information), the released barriers of $NO_2$ and NO are larger than that of $NH_3$.

**Origin of the $NO_3RR$ activity of promising catalysts.** Taking FeMo@g-CN and CrMo@g-CN as the representative catalysts, the results of differential charge densities reflect that the electrons are transferred from three O atoms to the dual-metal atoms. **Figure 3**e,f show the depletion of electrons occurring on O atoms while accumulation on dual-metal atoms, indicative of strong charge transfer between $NO_3^-$ and active site. Furthermore, charge transfer during $NO_3RR$ process on FeMo@g-CN and CrMo@g-CN (Figure S13, Supporting Information) is analyzed using the Bader charge method.

Each reactant-catalyst complex can be divided into three moieties for $NO_3RR$, which are g-CN substrate (moiety 1), $M_1M_2$ dimer (moiety 2) and adsorbates (moiety 3). As illustrated in Figure S13 (Supporting Information), it is found that the tendency of charge variation on FeMo@g-CN is analogous with that of CrMo@g-CN. Specially, the $M_1M_2$ dimers are positively charged (2.21-2.85 and 2.33-2.97 electrons transferred from FeMo and CrMo dimers, respectively) and reach the maximum value at the $*NH_3$ intermediate. The transferred charges of moiety 1 and 3 remain roughly unchanged during the steps of $*NO_3$-$*NO_3H$-$*NO_2$-$*NO_2H$-*NO-*NOH-*N, and gradually goes down (up) for moiety 1(3) during the steps of *N-*NH-* $NH_2$-$*NH_3$. We further calculate the minimum energy path (MEP) for the dissociation of FeMo, CrMo, CrRu and ZrMo dimers into two separated monomers to exclude the possibility of agglomeration and forming clusters. As shown in Figure S14, the diffusion barriers calculated by CI-NEB range from 4.75 to 12.32 eV. These results imply that four homogeneous DACs of FeMo@g-CN, CrMo@g-CN, CrRu@g-CN and ZrMo@g-CN have pretty high MEP with a superior kinetic stability.

To investigate the origin of the activation of $NO_3RR$, the couplings between $M_1M_2$ dimer *d* orbital and $NO_3^-$ orbital by investigating projected density of states (PDOS) (**Figure 4**). Electronic DOS of isolated $NO_3^-$ is shown in Figure S15 (Supporting Information), showing the anti-bonding states of $NO_3^-$ contain negligible *p* orbital. Upon the uptake of nitrate, the d band center of $M_1M_2$ dimer should shift upwardly, favoring the coupling with the antibonding orbital of $NO_3^-$. However, opposite tendency occurs in our study, suggesting that simple d band center can't explain the complex heterogeneous DACs. Recent researches have revealed that only one or two *d* sub-orbital participate in some catalytic reactions for SACs, namely orbital-dependent catalytic reactivity.[44-46] In order to explore whether this phenomena applies to heterogeneous DACs, we calculate the $dxy$, $dyz$, $dxz$, $dz^2$, $dx^2$-$y^2$ orbitals of $M_1M_2$ dimer in FeMo@g-CN and CrMo@g-CN (**Figure 4**e,f). It is revealed that all these *d* sub-orbitals couple with the anti-bonding orbitals ($\pi$*) of $NO_3^-$, indicating the indispensable role of each sub-orbital for the $NO_3RR$ activation. Similar trends are also found in CrRu@g-CN (Figure S16, Supporting Information) and ZrMo@g-CN (Figure S17, Supporting Information), indicating the activation of $NO_3RR$ originates from the synergistic effect of $M_1M_2$ dimer $dxy$, $dyz$, $dxz$, $dz^2$, $dx^2$-$y^2$ orbitals rather than some specific *d* sub-orbitals.

**Thermal stability and enhanced adsorption window of visible-light.** The thermal stability of catalytic material, especially for SACs and DACs, is an indispensable prerequisite because it may lead to collapse of the system and finally a degraded activity. Herein, AIMD simulations using canonical ensemble (NVT) at constant-temperature (400 K) in the Nosé-Hoover thermostat were performed to evaluate the thermal stability. At first, the velocities of atoms are randomly initialized by the Maxwell-Boltzmann distribution. Then, the systems are simulated for 10 ps with a timestep of 1 fs at the constant temperature. As evident in **Figure 5**a,b, the energy and temperature fluctuations of FeMo@g-CN and CrMo@g-CN oscillate near the equilibrium state, indicating the thermal stability with stable structure of these catalysts. Figures S18,19 (Supporting Information) show the energy and temperature fluctuations of some other selective $M_1M_2$@g-CN systems (TiCr@g-CN, TiZr@g-CN, CrRu@g-CN, , ZrMo@g-CN and ZrRu@g-CN). It can be seen that the thermal stability tends to deteriorate for TiZr@g-CN because the dual-metals tend to reconstruct and deviate from the original structure, which leads to the existence of two fluctuation curves (Figure S18b, Supporting Information) although a stable adsorption is inferred from the formation energy. This fact validates our previous work,[47] pointing that the criterion of stability is inadequate only by complying with of a favorable formation energy.

g-CN is well-known to possess good photoelectronic conversion to allow it a promising photocatalyst. To efficiently utilize the photons, a high absorption of wide spectrum of light is desired.[32] However, it is well-known that pristine g-CN can only adsorb ultraviolet (UV) light, which is also verified by the computation of optical adsorption spectra shown in **Figure 5**c,d. Interestingly, the deposition of metal dimers endows these DACs with anisotropic adsorption spectrum (**Figure 5**c,d), and broadens the adsorption window to capture the visible (VI) light. Figure S20 shows the band structures illustrating the variation of optical adsorption spectra of pristine g-CN, FeMo@g-CN, CrMo@g-CN, CrRu@g-CN and ZrMo@g-CN under the Heyd–Scuseria–Ernzerhof (HSE06) level. The bandgap of pristine g-CN monolayer is 3.23 eV (Figure S20a, Supporting Information), making them potentially absorbing the ultraviolet (UV) light. After the deposition of metal dimers, the FeMo@g-CN, CrMo@g-CN, CrRu@g-CN and ZrMo@g-CN are still semiconductors, and the bandgaps of them are reduced to 0.41, 0.52, 0.41 and 0.47 eV, respectively. The exciting changes endow these DACs to absorb the VI light. Therefore, compared with the pristine g-CN, these DACs can obtain a higher photoconversion efficiency as the efficient photocatalysts.

**Experimental prospect.** In nature, the $NO_3$RR process is realized through enzymatic cascades.[48] Firstly, the conversion of $NO_3^-$ to $NO_2^-$ could be catalyzed by the nitrate reductase Mo cofactor. Subsequently, the reduction of $NO_2^-$ to $NH_3$ is viable by the nitrate reductase Fe cofactor. This process implies that highly efficient $NO_3$RR may be realized by a catalytic cascade of bimetallic Fe-Mo sites. Inspired by the biological process, Murphy and co-workers synthesized atomically dispersed bimetallic FeMo-N-C monomer catalyst for the $NO_3$RR.[49] In their work, a synergized pathway was discovered that $NO_3^-$ is dissociated to $NO_2^-$ on Mo sites, while the subsequent reduction of $NO_2^-$ to $NH_3$ is achieved over Fe sites. Interestingly, previous theoretical work has revealed that Mo/g-CN and Fe/g-CN SACs with high PDS of 0.99 and 1.33 eV, implying the poor $NO_3$RR activity on single atom Mo- or Fe-decorated alone on g-CN.[8] Combined with previous experimental[49] and theoretical[8] works, our predicted FeMo dimer decorated on g-CN possessing the synergistic effect is expected to have highly efficient $NO_3$RR activity. Therefore, FeMo@g-CN that possesses the synergistic role of Mo-Fe metal dimer is worthy of further experimental synthesis and verification for $NO_3$RR. Similarly, CrMo@g-CN and CrRu@g-CN are also desirable for further experimental investigation. It should be noted that the experimentally synthesized FeMo-N-C catalyst[49] shows that Fe and Mo atoms are likely to keep isolated monomer forms separated by nitrogen and carbon atoms which is different from our FeMo dimmer embedded in CN network. Nevertheless, in both configurations, the Fe and Mo tends to form hybridized states with carbon and nitrogen, allowing a synergistic effect and activating the reduction reaction. In addition, similar synergistic effect may exist in other metal-CN hybridized systems.

For comparison the $NO_3RR$ activities for recent theoretical and experimental catalytic systems are illustrated in **Table 1**. The limiting potential of FeMo@g-CN is comparable to that of Ru/g-$C_3N_4$ (-0.34 V), $Ti_2$-Pc (-0.34 V), and $Cu_2$@NG, and exceeds that of Ti/g-CN (-0.39 V), Zr/g-CN (-0.41 V), Fe-$N_4$/C (-0.53 V), Os-$N_4$/C (-0.42 V), Pt-$N_4$C (-0.48 V), ZnSA-MNC (-0.58 V), and $Mn_2$-Pc (-0.41 V).

### Discussion

In summary, we put forth a feasible approach to boost $NO_3RR$ performance through a "donor-acceptor" couple of bimetallic active sites using comprehensive spin-polarized DFT calculations. Particularly, three $M_1M_2$@g-CN ($M_1M_2$ = FeMo, CrMo and CrRu) have been identified as the promising catalysts with high activity ($\Delta G_{max}$ < 0.5 eV), selectivity and stability for electrochemical $NO_3RR$. The PDS for these DACs is *NO + $H^+$ + $e^-$ → *NOH. In addition, the $NO_3RR$ activation originates from the synergistic effect of $M_1M_2$ dimer *d* orbital. We also show that the deposition of metal dimers can narrow the bandgaps of DACs, making them broaden the adsorption window of VI light for photocatalysis. Especially, our screened bimetallic FeMo@g-CN system as highly efficient electrochemical catalysts for $NO_3RR$ is consistent with latest experimental results on similar structure of FeMo-N-C system with efficient catalytic performance. This implies the effectiveness of FeMo dimers hybridized with carbon and nitrogen systems for synergistic activation of $NO_3RR$. This study points out that DACs with tunable composition of metallic active sites render great space for improving the activity and selectivity for electrochemical $NO_3RR$.

## Methods

**Computational details.** Spin-polarized DFT computations have been performed for structural relaxation within the Vienna *ab initio* simulation package (VASP).[51,52] The generalized gradient approximation (GGA) of Perdew-Burke-Ernzerhof (PBE) functional is employed to describe the exchange-correlation effects.[53] The ion-electron interaction is treated by the projector augmented wave (PAW) pseudopotentials,[54] whereas the kinetic cutoff energy is set to 450 eV. Single gamma-point sampling in Brillouin zone is used for structural optimization of 2 × 2 × 1 supercell, whereas a 3 × 3 × 1 grid within Monkhorst-Pack mesh is for self-consistent calculation. The required convergence accuracy will reach until the Hellman-Feynman forces acting on each ion are smaller than 0.02 eV/Å. A vacuum layer of at least 15 Å is employed to eliminate the interaction between adjacent images. The DFT+D3 method was used to describe the van der Waals (vdW) interactions.[55] The Hubbard U correction is not considered in this work due to the underestimation of U correction for adsorption energy.[8,56] Bader charge method is carried out to analyze the charge transfer.[57] Climbing image-nudged elastic band (CI-NEB) approach is used to calculate the energy barriers of gas desorption.[58] Since the $NO_3RR$ occurs in an aqueous environment, we also consider the solvent effect for proton transfer implemented by implicit solvation model VASPsol.[59] Heyd–Scuseria–Ernzerhof hybrid functional (HSE06) is used to accurately calculate the band structure and optical adsorption spectrum.[60]

The change of Gibbs free energies is calculated according to computational hydrogen electrode (CHE) model,[61] and the details are presented in Supplementary information. The most positive Gibbs free energy change ($\Delta G_{\mathrm{max}}$) in the entire electrochemical reaction is the potential-determining step (PDS). Thus, the limiting potential $U_{\mathrm{L}}$ is defined as the following equation:

$$U_{\mathrm{L}} = -\Delta G_{\mathrm{max}}/ne$$

where $n$ is the number of transferred electrons. To illustrate the thermal stability of the promising catalytic systems, the *ab initio* molecular dynamics (AIMD) simulations in the canonical ensemble (NVT) with constant temperature in the Nosé-Hoover heat bath were performed at 400 K for 10 ps.[62] During the simulations, the Brillouin-zone samplings adopt Γ-point grid.

The initial velocities of ions are randomly assigned by Maxwell-Boltzmann distribution at the targeting temperature. The time step is set to be 1 fs.

The d-band center can be described as[63]

$$\varepsilon_{\mathrm{d}} = \frac{\int_{-\infty}^{+\infty} n_{\mathrm{d}}(\varepsilon)\ \varepsilon\ d\varepsilon}{\int_{-\infty}^{+\infty} n_{\mathrm{d}}(\varepsilon)\ d\varepsilon}$$

## Acknowledgments

This work is supported by the Natural Science Foundation of China (Grant 22022309) and Natural Science Foundation of Guangdong Province, China (2021A1515010024), the University of Macau (MYRG2020-00075-IAPME) and the Science and Technology Development Fund from Macau SAR (FDCT-0163/2019/A3). This work was performed in part at the High-Performance Computing Cluster (HPCC) which is supported by Information and Communication Technology Office (ICTO) of the University of Macau.

## Conflict of Interest

The authors declare no conflict of interest.

## Author Contributions

Zheng Shu: Design, calculation and writing the original draft. Hongfei Chen, Xing Liu, Huaxian Jia and Hejin Yan: Analysis, discussion of the results and manuscript revision. Yongqing Cai: Supervision, funding acquisition and editing the manuscript.

## Data Availability Statement

The datasets generated and/or analyzed during the current study are available from the corresponding author upon reasonable request.

---

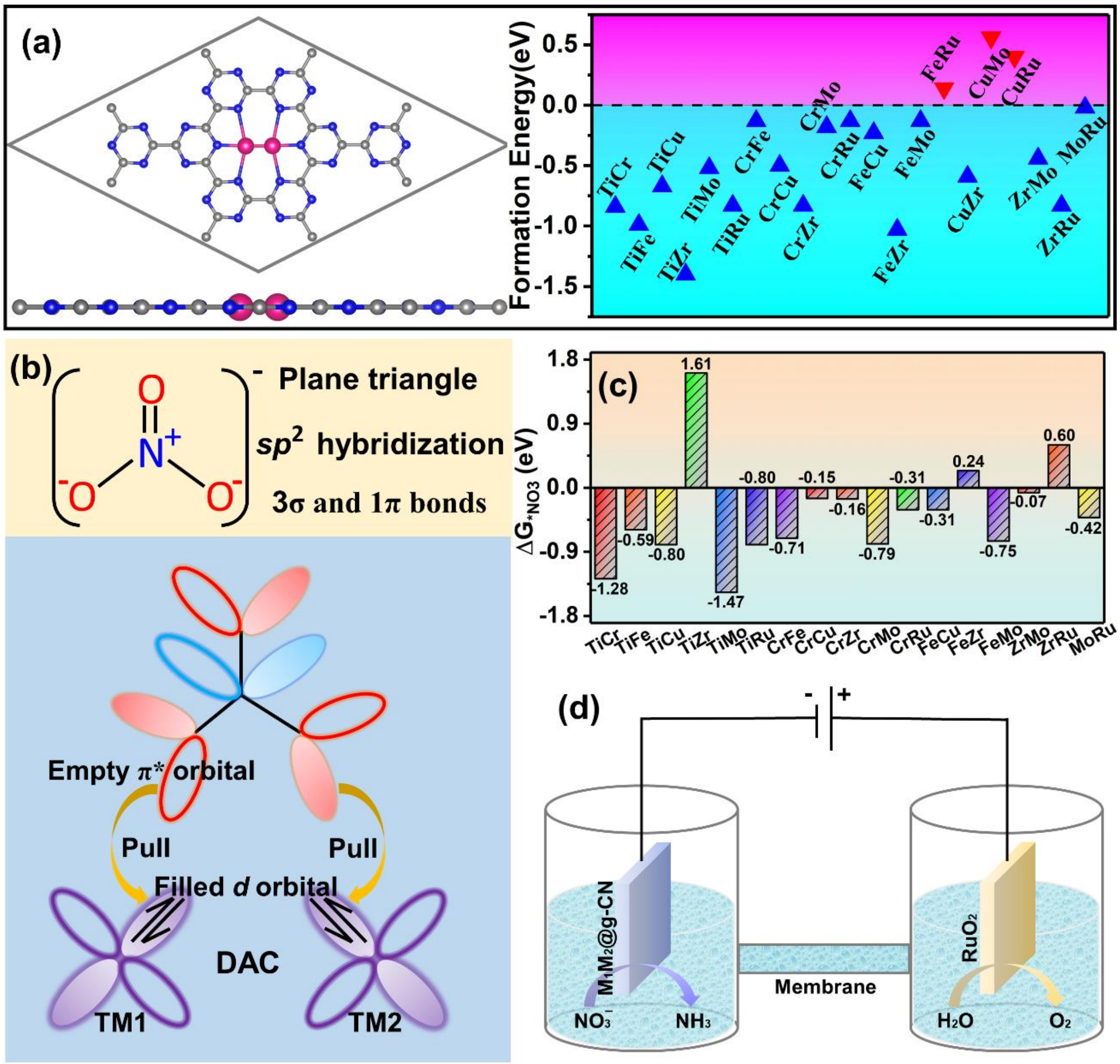

**Figure 1.** a) Atomic structure of $M_1M_2$@g-CN from top and side views and formation energies of various $M_1M_2$@g-CN substrates. Grey, blue and pink balls represent C, N atoms and $M_1M_2$ dimer, respectively. b) Overview of the activation mechanism on DACs for $NO_3$RR due to the anchoring of oxygen atoms with TM1 and TM2. c) The summary for the change of Gibbs free energy of $NO_3^-$ adsorption $\Delta G_{*NO3}$. d) Schematic diagram of the electrochemical cell for $NO_3$RR (cathode: $M_1M_2$@g-CN, anode: $RuO_2$ as an example).

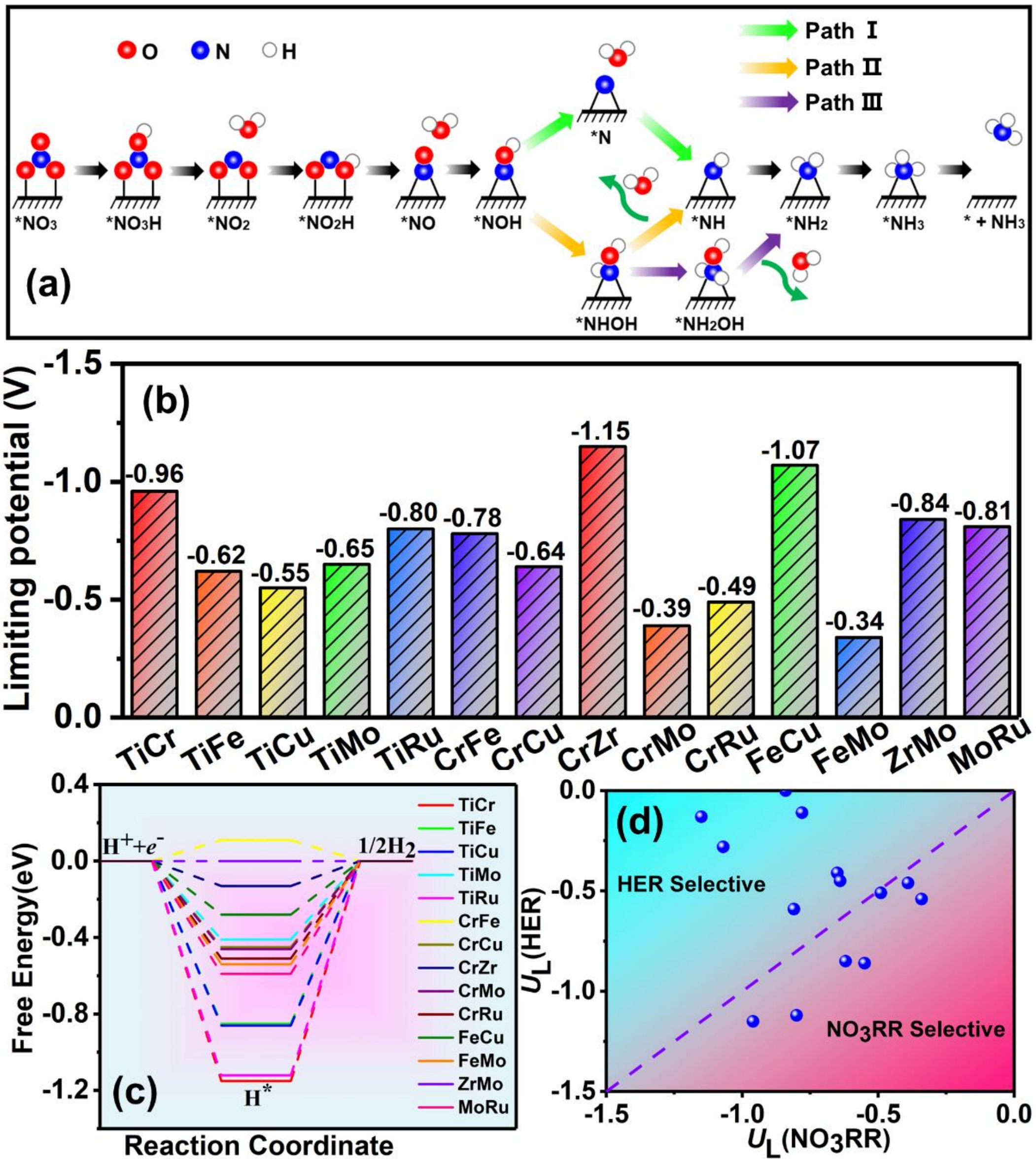

**Figure 2.** a) Pathways of $NO_3RR$, including Path I, II and III. b) Summary of limiting potentials on $M_1M_2$@g-CN for $NO_3RR$. c) Free energy diagrams of HER on $M_1M_2$@g-CN. d) Limiting potentials for $NO_3RR$ and HER illustrating the $NO_3RR$ selectivity of $M_1M_2$@g-CN.

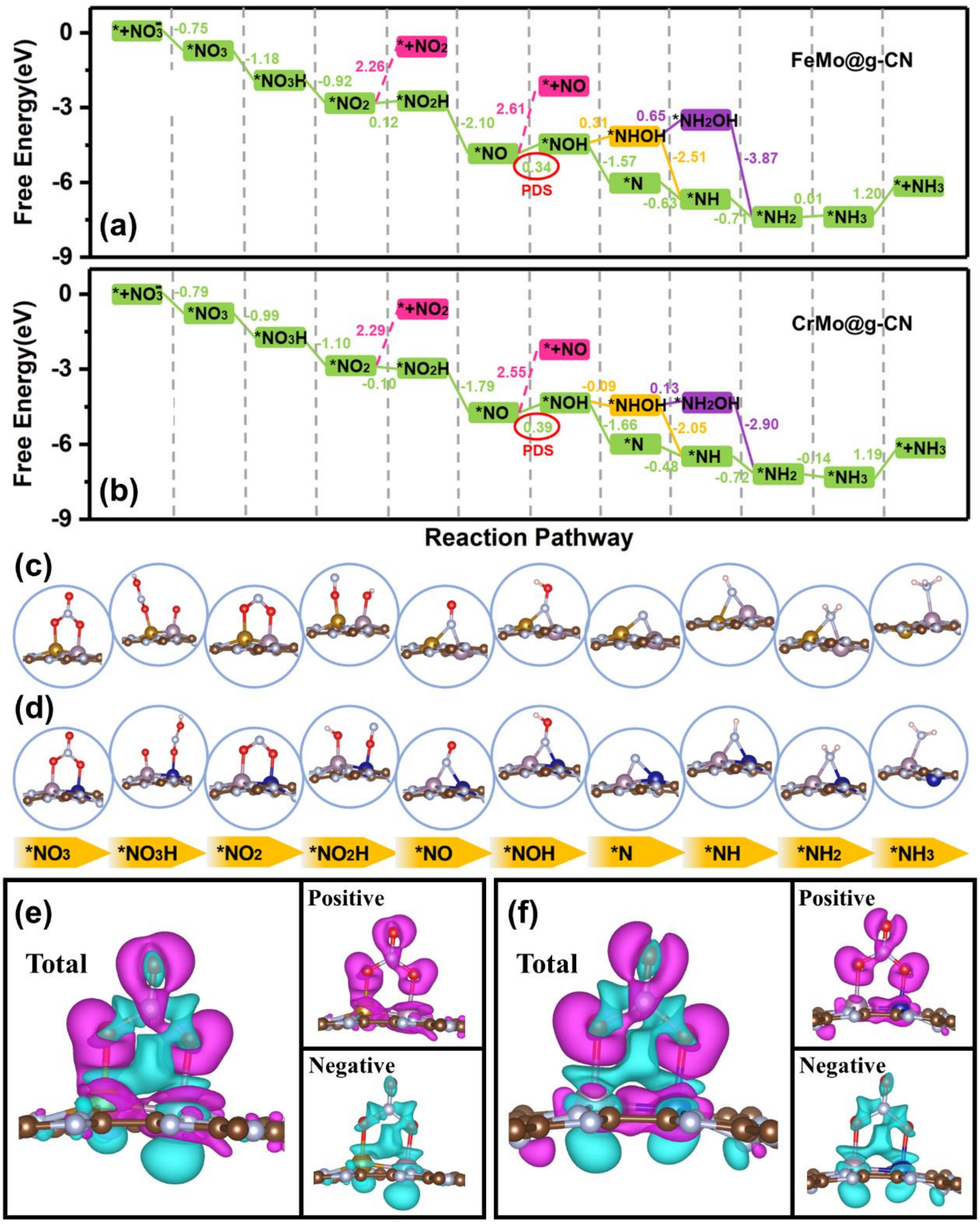

**Figure 3.** a, b) Free energy evolution of $NO_3RR$ along Path I (green), II (orange) and III (purple) on FeMo@g-CN and CrMo@g-CN, respectively. The energy injection of the release of $NO_2$ and NO are plotted for comparison. c, d) The optimized structures of $NO_3RR$ intermediates (along Path I) adsorbed on FeMo@g-CN and CrMo@g-CN, respectively. e, f) Iso-surface plot of differential charge density for $NO_3^-$ adsorbed at the dual-metal site of FeMo@g-CN and CrMo@g-CN, respectively. The iso-surface level is set to 0.0015 e $Å^{-3}$.

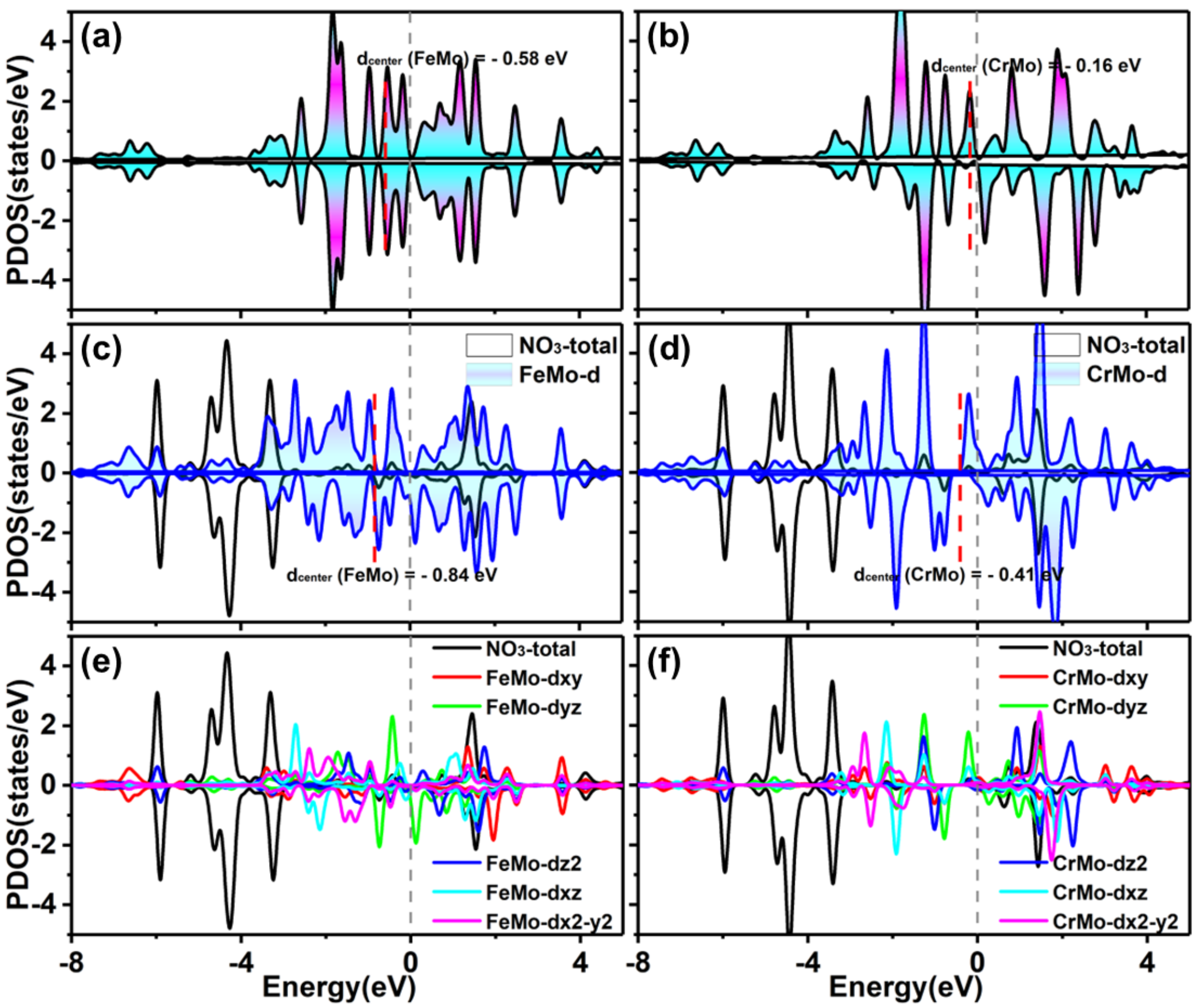

**Figure 4.** Projected density of states (PDOS) of dual-metal *d* orbital on a) FeMo@g-CN, b) CrMo@g-CN, c) $NO_3^-$ adsorbed FeMo@g-CN and d) $NO_3^-$ adsorbed CrMo@g-CN, respectively. PDOS of dual-metal *dxy*, *dyz*, *dxz*, $dz^2$, $dx^2$-$y^2$ orbitals on e) $NO_3^-$ adsorbed FeMo@g-CN and f) $NO_3^-$ adsorbed CrMo@g-CN, respectively. $E_F$ denotes the Fermi level, referring to 0 eV.

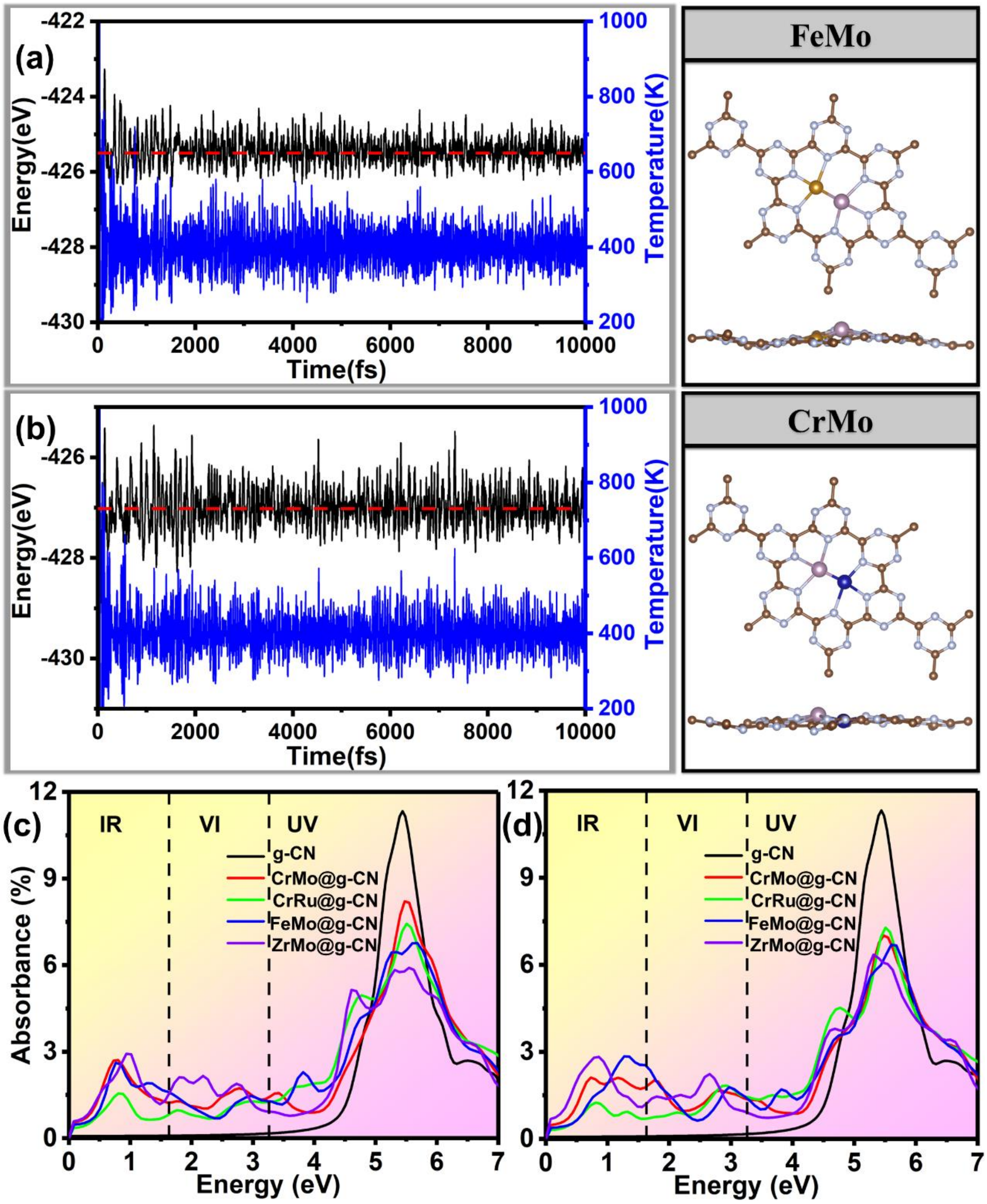

**Figure 5.** Fluctuations of energy and temperature with respect to time and their structures after AIMD simulations at 400 K of a) FeMo@g-CN and b) CrMo@g-CN, respectively. Optical adsorption spectra of FeMo@g-CN, CrMo@g-CN, CrRu@g-CN, ZrMo@g-CN and pure g-CN along the c) xx and d) yy axis.

**Table 1.** The comparison for $NO_3RR$ activity of this work and other catalytic systems.

| Catalyst | $\Delta G_{*NO3}$ (eV) | $U_L$ (V) | Potential-determining step | $\Delta G_{*NH3-des}$ (eV) | Reference |
|---|---|---|---|---|---|
| FeMo@g-CN | -0.75 | -0.34 | *NO → *NOH | 1.20 | this work |
| CrMo@g-CN | -0.79 | -0.39 | *NO → *NOH | 1.19 | this work |
| Ti/g-CN | -1.54 | -0.39 | *NO → *NOH | 0.96 | [8] |
| Zr/g-CN | -1.90 | -0.41 | $*NH_2 \rightarrow *NH_3$ | 1.08 | [8] |
| Ru/g-$C_3N_4$ | -2.44 | -0.34 | *NO → *NOH | 0.77 | [39] |
| Fe-$N_4$/C | -0.13 | -0.53 | *NO → *NOH | 0.34 | [26] |
| Os-$N_4$/C | -0.83 | -0.42 | *N → *NH | 0.01 | [26] |
| Pt-$N_4$/C | 0.73 | -0.48 | *NO → *NOH | -0.07 | [25] |
| ZnSA-MNC | 0.16 | -0.58 | $*NO_3 \rightarrow *NO_3H$ | 0.35 | [27] |
| $Ti_2$-Pc | -4.04 | -0.34 | $*O_2H_3 \rightarrow 2H_2O$ | | [30] |
| $Mn_2$-Pc | -0.55 | -0.41 | $*NO_3 \rightarrow *NO_3H$ | | [30] |
| $Cu_2$@NG | -0.73 | -0.36 | $*NO_3 \rightarrow *NO_3H$ | -0.04 | [50] |